\RequirePackage{fix-cm}
\documentclass[pdftex,twocolumn,final,epjc3]{svjour3}  

\usepackage{xspace}
\usepackage{enumitem}
\usepackage{amsmath}
\usepackage{amssymb}
\usepackage{nicefrac}
\usepackage{xcolor}
\usepackage{graphicx}
\usepackage{hyperref}
\usepackage{cite}
\usepackage{wasysym}
\usepackage[frozencache]{minted}

\usepackage[normalem]{ulem}

\usepackage{cuted}
\usepackage{etoolbox}
\AtBeginEnvironment{strip}{\leavevmode}

\smartqed
\journalname{Eur. Phys. J. C}

\newcommand{\xBj}{x_{\mathrm{Bj}}}

\newcommand{\eg}{\textit{e.g.}\xspace}
\newcommand{\ie}{\textit{i.e.}\xspace}

\newcommand{\etc}{\textit{etc.}\xspace}

\def\intd{\mathrm{d}}

\newminted{cpp}{breaklines=true,fontsize=\footnotesize,texcomments=true}
\newminted{bash}{breaklines=true,fontsize=\footnotesize,texcomments=true}
\newminted{xml}{breaklines=true,fontsize=\footnotesize,texcomments=true}

\newmintinline{cpp}{breaklines=true}
\newmintinline{bash}{breaklines=true}
\newmintinline{xml}{breaklines=true}

\begin{document}

\title{EpIC: novel Monte Carlo generator for exclusive processes}

\author{
E.~C.~Aschenauer\thanksref{emailEA,addressBNL}
\and
V.~Batozskaya\thanksref{emailVB,addressNCBJ}
\and
S.~Fazio\thanksref{emailSF,addressCAL}
\and
K.~Gates\thanksref{emailKG,addressGlasgow}
\and
H.~Moutarde\thanksref{emailHM,addressCEA}
\and
D.~Sokhan\thanksref{emailDS,addressCEA,addressGlasgow}
\and
H.~Spiesberger\thanksref{emailHS,addressMainz}
\and
P.~Sznajder\thanksref{emailPS,addressNCBJ}
\and
K.~Tezgin\thanksref{emailKT,addressBNL}
}

\thankstext{emailEA}{e-mail: elke@bnl.gov}
\thankstext{emailVB}{e-mail: varvara.batozskaya@ncbj.gov.pl}
\thankstext{emailSF}{e-mail: salvatore.fazio@unical.it}
\thankstext{emailKG}{e-mail: k.gates.1@research.gla.ac.uk}
\thankstext{emailHM}{e-mail: herve.moutarde@cea.fr}
\thankstext{emailDS}{e-mail: daria.sokhan@cea.fr}
\thankstext{emailHS}{e-mail: spiesber@uni-mainz.de}
\thankstext{emailPS}{e-mail: pawel.sznajder@ncbj.gov.pl}
\thankstext{emailKT}{e-mail: ktezgin@bnl.gov}

\institute{
Department of Physics, Brookhaven National Laboratory, Upton, New York 11973 \label{addressBNL}
\and
~National Centre for Nuclear Research (NCBJ), Pasteura 7, 02-093 Warsaw, Poland \label{addressNCBJ}
\and
~University of Calabria \& INFN-Cosenza, Italy \label{addressCAL}
\and
~University of Glasgow, Glasgow G12 8QQ, United Kingdom \label{addressGlasgow}
\and
~IRFU, CEA, Universit\'e Paris-Saclay, F-91191 Gif-sur-Yvette, France \label{addressCEA}
\and
~PRISMA$^{+}$ Cluster of Excellence, Institut f{\"u}r Physik,
Johannes Gutenberg-Universit{\"a}t, D-55099 Mainz, Germany\label{addressMainz}
}

\date{Received: date / Accepted: date}

\maketitle

\sloppy

\begin{abstract}
We present the EpIC Monte Carlo event generator for exclusive processes sensitive to generalised parton distributions. EpIC utilises the PARTONS framework, which provides a flexible software architecture and a variety of modelling options for the partonic description of the nucleon. The generator offers a comprehensive set of features, including multi-channel capabilities and radiative corrections. It may be used both in analyses of experimental data, as well as in impact studies, especially for future electron-ion colliders. 
\end{abstract}
\section{Introduction}
\label{sec:intro}

The objective of this study is to develop a Monte Carlo event generator for exclusive processes involving hadrons remaining coherent during interactions with high-energy leptons. The factorisation theorems developed in the framework of quantum chromodynamics (QCD) allow one to describe such processes in terms of non-perturbative generalised parton distribution functions (GPDs) \cite{Muller:1994ses, Ji:1996ek, Ji:1996nm, Radyushkin:1996ru, Radyushkin:1997ki} convoluted with perturbatively calculable hard scattering subprocesses. 

GPDs are universal, process-independent functions that parametrise the off-forward nucleon matrix elements of quark and gluon bilinear operators with light-like separations. In case there is no momentum transfer to the nucleon, \ie in the forward limit, certain GPDs become equivalent to PDFs. Additionally, the first Mellin moments of GPDs are related to elastic form factors. In this regard, GPDs may be viewed as a unified concept of elastic form factors studied via elastic scattering processes and one-dimensional parton distribution functions studied via (semi-) inclusive scattering processes. Another key aspect of GPDs is their relation to nucleon tomography. The Fourier transform of GPDs are related to the impact parameter space distributions when there is no collinear, but finite transverse momentum transfer to the nucleon \cite{Burkardt:2000za, Burkardt:2002hr, Burkardt:2004bv}. This relation enables nucleon tomography, which involves the correlation between impact space distribution functions, parton polarization, and the longitudinal momentum fractions carried by partons. GPDs also exhibit a unique relationship with energy-momentum tensor (EMT) form factors, which encode fundamental properties of the nucleon, such as mass and spin decomposition of the nucleon into its constituent parts (including orbital angular momentum) \cite{Ji:1996ek, Ji:1996nm, Ji:1997gm} as well as its internal structure based on the so-called ``mechanical’’ forces \cite{Polyakov:2002yz, Polyakov:2018zvc, Lorce:2018egm}. For more information on GPDs, we refer to the available reviews on the subject, like Ref. \cite{Diehl:2003ny}.  

The extraction of GPDs from data on exclusive processes is not an easy task, mainly due to the difficulty of the inverse problem one must solve. Namely, typically many types of GPDs contribute to a given process, and it is necessary to deconvolute them all from the process amplitude. In order to accomplish this, measurements of many different types of processes are needed in a wide range of kinematic domains. Therefore, measurements of exclusive processes have been performed in several facilities, such as DESY, JLAB, and CERN. In addition to that, GPDs are pillars of scientific programmes for a new generation of machines. This includes both electron-ion colliders, such as EIC \cite{Accardi:2012qut, AbdulKhalek:2021gbh}, EIcC \cite{Anderle:2021wcy} and LHeC \cite{LHeCStudyGroup:2012zhm}, and fixed target experiments, such as AMBER at CERN \cite{Adams:2676885} and JLAB12 \cite{Arrington:2021alx}.

\begin{figure}[!ht]
\begin{center}
\includegraphics[width=0.3\textwidth]{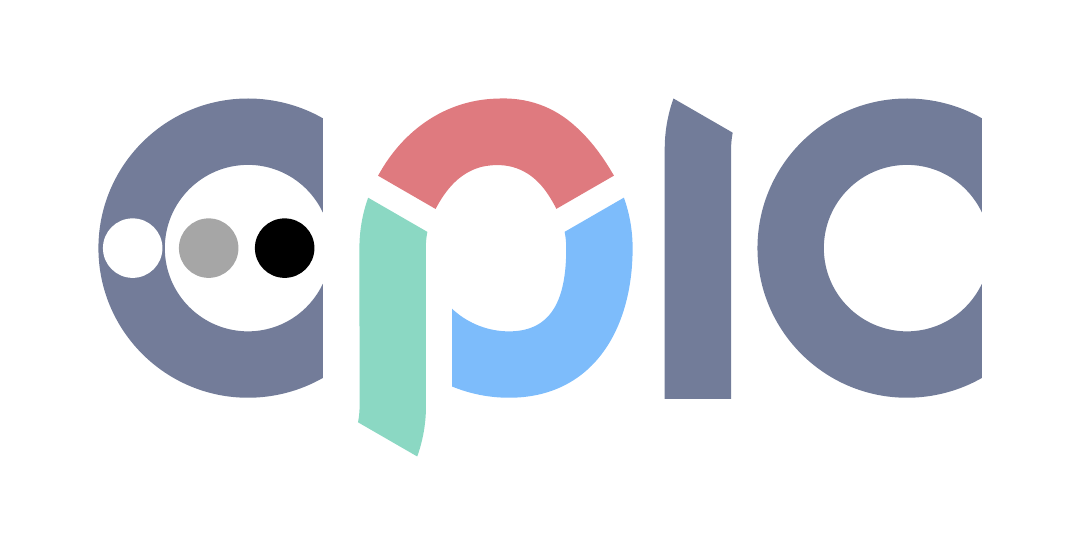}
\caption{Logo of the project.}
\label{fig:intro:logo}
\end{center}
\end{figure}

As part of this global effort, we have developed a Monte Carlo (MC) generator called EpIC, whose logo appears in Fig. \ref{fig:intro:logo}. The EpIC generator features a novel architecture based on the modularity programming paradigm. In this way, the code structure is kept as simple as possible and the addition of new developments, such as new channels or algorithms to generate random numbers, is made as easy as possible. The architecture of EpIC, including the nomenclature for naming its elements, is based on the PARTONS framework \cite{Berthou:2015oaw}. PARTONS is also used to evaluate the Born cross-section for a given process, which is used to generate MC events after the inclusion of radiative corrections (RCs).

EpIC can generate events for a number of exclusive processes. The following are available at the time of this writing: deeply virtual Compton scattering (DVCS), time-like Compton scattering (TCS), and deeply virtual meson production (DVMP) of $\pi^0$ and $\pi^+$ mesons. In this article, we choose the case of DVCS whenever an exemplary process is needed. Adapting to other processes is straightforward and typically only requires replacing the acronym, \eg the equivalence of \texttt{DVCSGeneratorService} for TCS is \texttt{TCSGeneratorService}. EpIC, as in the case of PARTONS, follows the so-called Trento convention \cite{Bacchetta:2004jz} for the definition of \eg scattering angles.

The purpose of this article is to provide a reference to be used for a better understanding of EpIC's structure and its functioning. We will mostly focus on the explanation of the architecture, starting with a brief description of the PARTONS framework in Sect. \ref{sec:partons}. Topics covered in that section are crucial to understand the technical aspects of EpIC. In Sect. \ref{sec:arch}, we present the architecture of EpIC. In particular, we list all types of modules and show the relation between them. The user interface is introduced in Sect. \ref{sec:ui}, while the demonstration of the generator's performance is shown in Sect. \ref{sec:demo}. The summary is given in Sect. \ref{sec:summary}. In the appendix of the article (\ref{sec:rc}), we review the basics of radiative corrections and explain how to compute them in the collinear approximation. The purpose of this appendix is to help users familiarise themselves with the parameters of RC used in EpIC. 

EpIC is written in the C++ programming language. The code can be accessed at the GitHub platform \cite{www:epic_github} and is distributed under the GPL 3.0 licence. The project's webpage, Ref. \cite{www:epic},  includes information such as a technical documentation of C++ classes, a guide on how to compile the code, and example code for users.

\section{PARTONS framework}
\label{sec:partons}

PARTONS (PARtonic Tomography Of Nucleon Software) \cite{Berthou:2015oaw} is a software framework for studying the 3D structure of hadrons. It provides essential tools for QCD phenomenology and allows one to compute observables from the models of non-perturbative objects. For instance, it includes various models of GPDs and methods to perform their pQCD evolution. It also includes methods to evaluate amplitudes and cross-sections for a variety of exclusive processes, like DVCS and TCS, described at various orders of pQCD precision. The framework can be used in analyses to predict observables from existing models, see \eg Refs. \cite{Grocholski:2019pqj,Anderle:2021wcy,AbdulKhalek:2021gbh}, but also to constrain new models from available experimental data \cite{Moutarde:2018kwr,Moutarde:2019tqa}. Because of its versatility, the framework can also act as a laboratory for studying new concepts for phenomenology, like new ways of modelling, see \eg Refs. \cite{Bertone:2021yyz,Chavez:2021llq,Dutrieux:2021wll}. 

What distinguishes PARTONS from other software projects used in the field of particle physics is its architecture utilising the modular programming paradigm. The central role in this architecture, which has been also adopted in EpIC, is played by \emph{the modules}. A module is a single encapsulated development of a given type, for instance a single GPD model. The only purpose of modules is to provide data for a specific input, like values of GPDs for a given GPD kinematics. The construction of modules utilises both the inheritance and polymorphism mechanisms of C++. Therefore, the development of new modules is remarkably straightforward, as developers need only to provide their code related to the developments they are interested in and adapt it to the predefined classes in the framework. We demonstrate this with the example of \cppinline{GPDGK11} module, which is used to implement the Goloskokov-Kroll (GK) GPD model \cite{Goloskokov:2005sd,Goloskokov:2007nt,Goloskokov:2009ia}, from the PARTONS library. The following is an excerpt from the header file:
\begin{cppcode}
namespace PARTONS {

class GPDGK11: public GPDModule {

public:

//Unique ID to automatically register the class in the registry.
static const unsigned int classId; 
	
//Constructor.
//Class name is passed to GPDModule constructor.
GPDGK11(const std::string& className);

//Used for automation.
virtual void configure(const ElemUtils::Parameters& parameters);

//Function containing the evaluation of GPD $H$. 
virtual PartonDistribution computeH();

...
};
}
\end{cppcode}
where only the most relevant members of the class are shown, some of which will be discussed further in this section. The purpose of the \cppinline{GPDGK11} module is to evaluate the values of GPDs. This task is predefined in \cppinline{GPDModule}, being in terms of inheritance the parent class of \cppinline{GPDGK11}. The predefinition means here the declaration of a virtual function, so that each derived classes (\cppinline{GPDGK11}, \cppinline{GPDGK16}, \dots) have a function of the same signature among its members, but the actual implementation of this function in those classes is in general different. The evaluation of GPDs takes place, for instance, in the following function:
\begin{cppcode}
using namespace PARTONS;

PartonDistribution GPDGK11::computeH(){

//Protected variables that store input GPD kinematics to be used by developers. 
m_x; m_xi; m_t; m_MuF2, m_MuR2;

...

//Container to be returned. 
PartonDistribution partonDistribution;

//Store values of $H$, $H^{(+)}$ and $H^{(-)}$, respectively, for up quarks.
//Results for other flavours can be added analogously. 
partonDistribution.addQuarkDistribution( QuarkDistribution( QuarkFlavor::UP, ..., ..., ...)); 

//Store value of $H$ for gluons.
partonDistribution.setGluonDistribution(...);

return partonDistribution;
}
\end{cppcode}
The function \cppinline{GPDGK11::computeH()} returns a container that stores the numerical values of GPDs of type $H$, including singlet, $H^{(+)}$, and non-singlet, $H^{(-)}$, combinations, as defined in Ref. \cite{Diehl:2003ny}. The input GPD kinematics is accessible in the body of the function via the \cppinline{m_x}, \cppinline{m_xi}, \cppinline{m_t}, \cppinline{m_MuF2} and \cppinline{m_MuR2} variables, which are defined in the \cppinline{GPDModule} class. Those variables correspond to $x$ (average longitudinal momentum fraction carried by the active parton), $\xi$ (skewness parameter), $t$ (four-momentum transfer to the hadron target), $\mu_\mathrm{F}^2$ (factorisation scale squared) and $\mu_\mathrm{R}^2$ (renormalisation scale squared) GPDs depend on. They are set when one evaluates GPDs via \cppinline{PartonDistribution compute(const GPDKinematic& kinematic, GPDType::Type gpdType)}, which is also defined in \cppinline{GPDModule}. As one may see, in our example quark flavours and GPD types are distinguished by \cppinline{QuarkFlavor} and \cppinline{GPDType} enums, respectively, while inputs and outputs are encapsulated in the \cppinline{QuarkDistribution}, \cppinline{PartonDistribution} and \cppinline{GPDKinematic} containers.

A single instance of each module is loaded into memory at the beginning of the program execution, namely, when static constant variables are initialised. The addresses of such instances are stored by \emph{the registry}, which later acts as a phone book, allowing one to access the instances by either their IDs or defined names. We demonstrate this mechanism with the following example:
\begin{cppcode}
using namespace PARTONS;

const unsigned int GPDGK11::classId = BaseObjectRegistry::getInstance()-> registerBaseObject(new GPDGK11("GPDGK11"));
\end{cppcode}
When the program starts, \cppinline{GPDGK11::classId} is initialised with a unique ID assigned by the registry (\cppinline{BaseObjectRegistry} class being a singleton). During this process, the registry takes and stores the address of each new instance. All modules inherit from the same basic class called \cppinline{BaseObject}, allowing the registry to store addresses of the same type of objects. The constructor of \cppinline{BaseObject} requires a unique string of characters that is used as a human-readable name of the module.

The registry stores addresses of single instances of pre-configured modules. Users may get copies of those instances by using \emph{the factory} (also being a singleton), for instance:
\begin{cppcode}
using namespace PARTONS;

GPDModule* pGPDModel = nullptr;

//Get by ID.
pGPDModel = Partons::getInstance()-> getModuleObjectFactory()-> newGPDModule(GPDGK11::classId);

//Get by name
pGPDModel = Partons::getInstance()-> getModuleObjectFactory()-> newGPDModule("GPDGK11");
\end{cppcode}
The registry and factory are the key architectural ingredients, which allow users to add an unlimited number of new modules in the PARTONS architecture, without having to modify existing modules.

There are many types of modules, but one may \emph{configure} all of them, \ie pass additional information to the modules, in a generic way, avoiding, for instance, the static casting. This is possible thanks to the \cppinline{void configure(const ElemUtils::Parameters& parameters)} virtual functions, which developers may use to allow  transferring additional parameters to modules. For instance:
\begin{cppcode}
using namespace PARTONS;

void configure(const ElemUtils::Parameters& parameters){

  //Check if parameter is available.
  if (parameters.isAvailable("myKey")) {
  
    //Set.
    int myVariable = parameters.getLastAvailable().toInt();
  }
}
\end{cppcode}
can be used to set \cppinline{myVariable} variable to $1$ in the following way:
\begin{cppcode}
//Container for multiple parameters.
ElemUtils::Parameters parameters;

//Set.
parameters.add(ElemUtils::Parameter("myKey", 1));

//Pass.
pMyModule->configure(parameters);
\end{cppcode}
where \cppinline{"myKey"} is the unique name of the parameter. The way of supplying modules with additional parameters presented here plays a crucial role in constructing user interface (UI).

To avoid ``manually’’ repeating some tasks, \emph{the services} are used. In this way, users can perform complex tasks in a straightforward and robust manner by hiding the complexity of low-level functions. The services link such tasks as parsing input files, running many computations, employing multi-threading computing, filling the database, printing the output to the screen \etc Without services, all those tasks would have to be done explicitly by users, requiring a profound knowledge of the functioning of the project. This way of working would also be time consuming and vulnerable to mistakes. 

\section{EpIC's architecture}
\label{sec:arch}

\begin{figure*}[!ht]
\begin{center}
\includegraphics[width=1.0\textwidth]{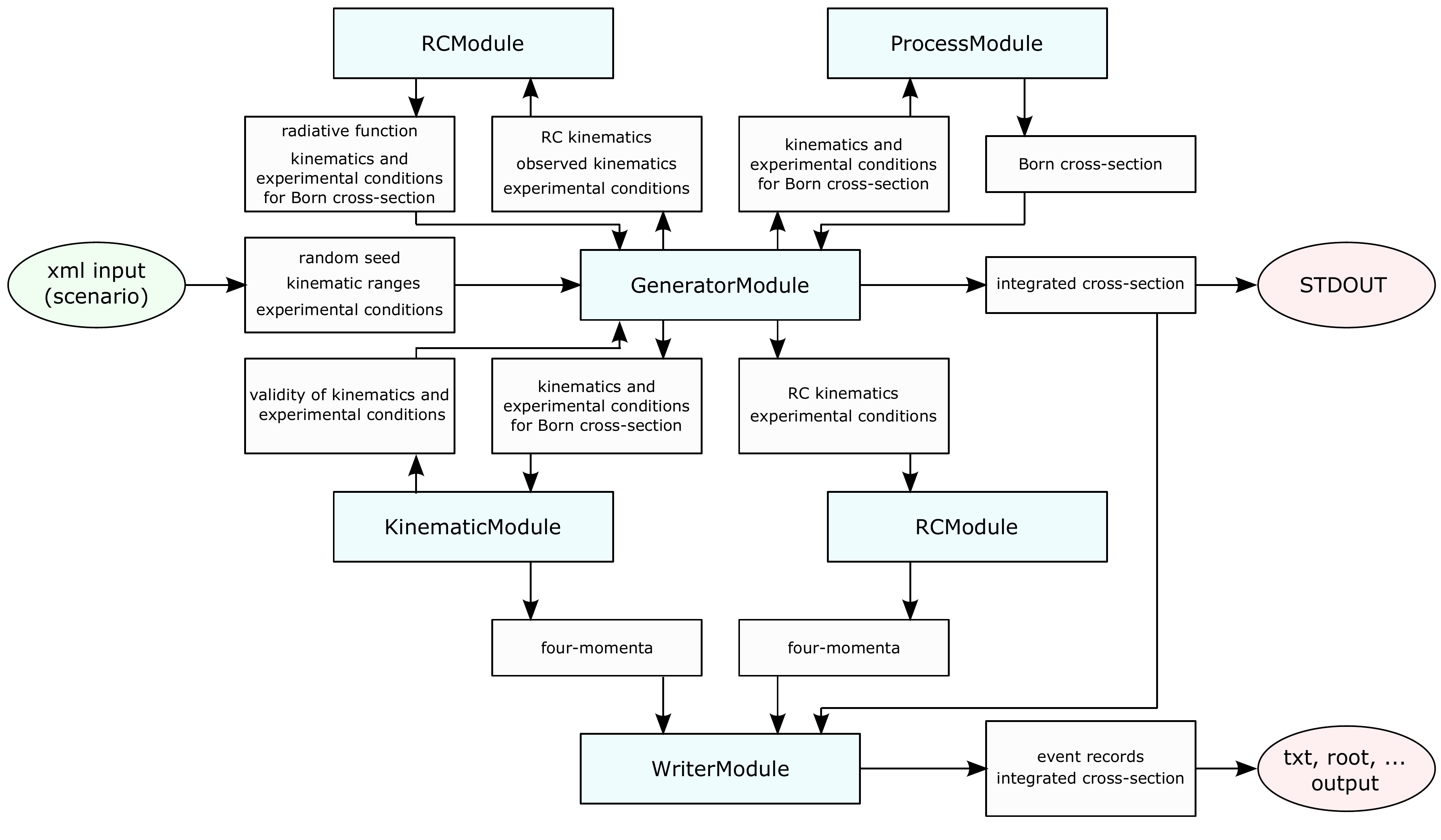}
\caption{Sketch showing the generic architecture of the project, \ie without specifying the corresponding physics process for some modules. Details are provided in the text.}
\label{fig:arch:arch}
\end{center}
\end{figure*}

\begin{figure}[!ht]
\begin{center}
\includegraphics[width=0.48\textwidth]{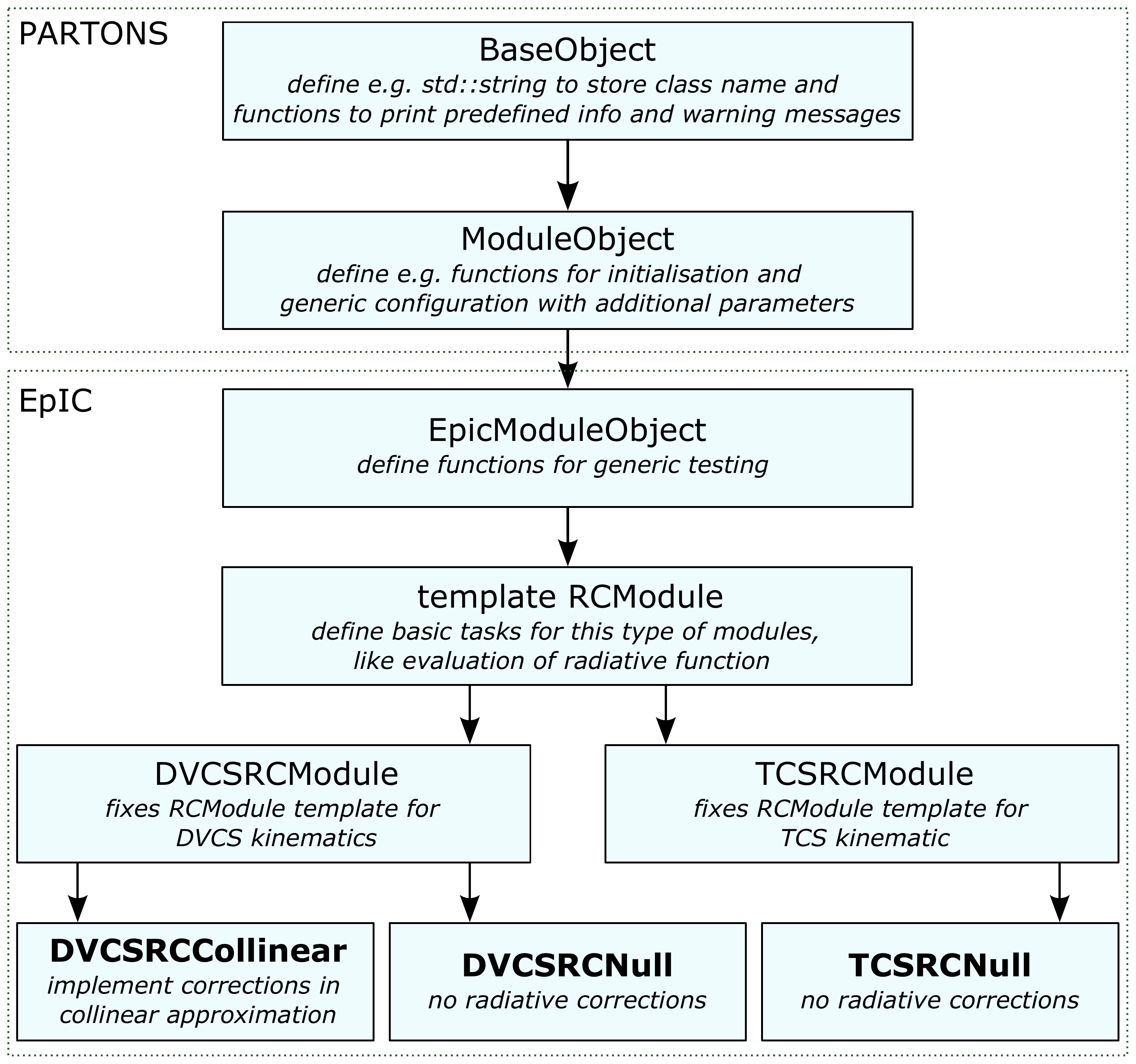}
\caption{Inheritance tree of modules used for the implementation of radiative corrections. Each box represents a single C++ class. Two first classes are the part of PARTONS library, the rest are the part of EpIC project. Classes acting as modules to be used be users are indicated by boldface fonts.}
\label{fig:arch:inh}
\end{center}
\end{figure}

The generic architecture of the generator, including the flow of data transferred between the modules, is shown in Fig.~\ref{fig:arch:arch}. Each module takes input data, processes it according to predefined tasks, and returns the corresponding output. The actual implementation of some types of modules depends on the simulated process. For instance, \cppinline{RCModule} is the parent class for \cppinline{DVCSRCModule} and \cppinline{TCSRCModule}, which are used for DVCS and TCS processes, respectively. These two types of modules process data according to the same task predefined in \cppinline{RCModule}, but they use different inputs. It is either DVCS or TCS kinematics. Because of this difference, \cppinline{RCModule} has been developed as a C++ template, allowing \cppinline{DVCSRCModule} and \cppinline{TCSRCModule} to handle input containers of various types. With those templates, the addition of new processes to the generator is as easy as possible. \cppinline{DVCSRCModule} and \cppinline{TCSRCModule} are used as base classes for actual modules, where specific physics developments are encoded. In Fig.~\ref{fig:arch:inh} we present the entire tree of inheritance for modules of \cppinline{RCModule} type. 

Below, we provide a brief description of each type of module and its purpose. Additionally, we indicate whether the actual implementation of modules of a given type depends on the simulated process. This information is useful in recognising which parts of the generator need to be extended when a new process is added to the generator.

\subsection{RCModule} 
\noindent\emph{Purpose:} Simulation of radiative corrections.

\smallskip
 
\noindent\emph{Does it depend on the generated process?} Yes. 

\smallskip

\noindent\emph{Description:} In EpIC events are generated according to the probability distribution given by: 
\begin{equation}
\mathrm{d}\sigma(X', Z) = \mathrm{d}\sigma_{0}(X) ~ R(Z)\,.
\label{eq:arch:BornTimesRC}
\end{equation}
Here, $d\sigma_{0}(X)$ is the Born cross-section (given by PARTONS), while $R(Z)$ is the radiator function. The set of variables that the Born cross-section depends on is denoted by $X$, for instance $X = \{\xBj, t, Q^2, \phi, \phi_{S}, E_{l} \}$ for DVCS, where $\xBj$ and $Q^2$ are the usual DIS variables, $\phi$ is the angle between the lepton scattering plane and the production plane, $\phi_{S}$ is the angle between the lepton scattering plane and the target spin component perpendicular to the direction of the virtual photon, and $E_{l}$ is the lepton energy in the fixed target frame \cite{Berthou:2015oaw}. Because of the radiative corrections, the values of those variables may be different from those for a generated event, $X'$. The radiator function, on the other hand, depends on the set of variables $Z$. The number of variables making this set depends on the approximation one uses. For instance, if one considers the collinear approximation and only the initial and final state radiations of single photons from the incoming and scattered leptons, two variables make the set $Z$, namely $Z = \{z_{1}, z_{3}\}$, see \ref{sec:rc} for more details. 

Taking the above in mind, modules of \cppinline{RCModule} type perform three actions: \emph{i}) they define a number of variables that the radiative corrections depend on, \ie they define the set $Z$, \emph{ii}) they implement a function used for the evaluation of $R(Z)$, but also for the evaluation of ``true'' kinematics entering the Born cross-section, $X = f(X', Z)$, \emph{iii}) they implement a function that stores the four-momenta of radiated photons in the final event records.

\subsection{ProcessModule} 
\noindent\emph{Purpose:} Evaluation of Born cross-section.

\smallskip
 
\noindent\emph{Does it depend on the generated process?} Yes. 

\smallskip

\noindent\emph{Description:} The Born cross-section is evaluated in EpIC using the modules of \cppinline{ProcessModule} type, which are defined in the PARTONS library. In the most general case, where the Born cross-section is directly calculated from GPDs, \cppinline{ProcessModule} requires modules of other types for proper functioning: \cppinline{ConvolCoeffFunctionModule} for the evaluation of the process amplitudes, \cppinline{GPDModule} for the evaluation of GPDs at a reference scale, \cppinline{GPDEvolutionModule} for the evolution of GPDs, \cppinline{XiConverterModule} for the evaluation of GPD skewness variable, $\xi$, from the process kinematics, and \cppinline{ScalesModule} for the evaluation of renormalisation and factorisation scales from the process kinematics. However, this general case can hardly be used in the Monte Carlo generation because of the nested integrations evaluating the Born cross-section. Instead, one may start the evaluation, for instance, from the level of amplitudes parametrised in lookup tables.  In this case, \cppinline{GPDModule} and \cppinline{GPDEvolutionModule} are not used as there is no need for a convolution of GPDs with hard scattering coefficient functions. More details about the \cppinline{ProcessModule} modules can be found in Ref. \cite{Berthou:2015oaw}. 

\subsection{EventGeneratorModule} 
\noindent\emph{Purpose:} Generation of kinematic configurations according to the probability distribution given by the product of the Born cross-section and the radiator function.

\smallskip
 
\noindent\emph{Does it depend on the generated process?} No.

\smallskip

\noindent\emph{Description:} Modules of this type play a central role in the generation of events. Three actions are implemented in those modules: \emph{i}) Initialisation, where the algorithm implemented in the module probes the probability distribution, for instance to ``memorise’’ it using an interpolation method, or to capture its key features, like the location of places in which the probability distribution tends to change rapidly. For some complicated probability distribution functions, the initialisation process can be time-consuming, but the results of the process can be stored in a file for use in subsequent jobs. \emph{ii}) Generation of kinematic configurations according to the probability distribution, based on the data collected during the initialisation stage. \emph{iii})  Evaluation of the total cross-section, which is needed to normalise the event distributions to a given integrated luminosity.

\subsection{KinematicModule} 
\noindent\emph{Purpose:} Evaluation of four-momenta for a given kinematic configuration.

\smallskip
 
\noindent\emph{Does it depend on the generated process?} Yes. 

\smallskip

\noindent\emph{Description:} Modules of this type are responsible for the evaluation of four-momenta for a given kinematic configuration. In addition, they implement a predefined function that determines if a given kinematic configuration is physical, \ie if it does not break any kinematic limit. This function is utilised by \cppinline{GeneratorModule} to probe only valid kinematics.

\subsection{WriterModule} 
\noindent\emph{Purpose:} Composing and saving event records. 

\smallskip
 
\noindent\emph{Does it depend on the generated process?} No.

\smallskip

\noindent\emph{Description:} The purpose of modules of this type is to create and store event records in output files. Different formats may be implemented by different modules. Additional information, like the  integrated cross-section, can also be stored in output files, \eg in their headers. 
\section{User interface}
\label{sec:ui}

Both compilation and linking of the project make use of the CMake tool \cite{www:cmake}. In the current version ($1.0.0$) EpIC requires the following external libraries: PARTONS \cite{Berthou:2015oaw} (providing elements of the architecture and used for the evaluation of Born cross-sections), ROOT \cite{Brun:1997pa} (used in one of the \texttt{EventGeneratorModule} modules), HepMC3 \cite{Buckley:2019xhk} (used in one of the \texttt{WriterModule} modules), GSL \cite{galassi2009gnu} (used for the generation of random numbers). For more detailed and always up-to-date information we refer to the online documentation \cite{www:epic}.  
 
The executable of the project, \bashinline{epic}, must be invoked with two arguments, like this: 
\begin{bashcode}
./epic --seed=SEED --scenario=SCENARIO_PATH
\end{bashcode}
where \texttt{SEED} is the random seed (unsigned integer) to be used in the initialisation of modules that deal with random numbers, and \texttt{SCENARIO\_PATH} is the relative or absolute path to \emph{the scenario} containing all options used in the generation. 

The EpIC scenarios are written in XML markup language which, thanks to the usage of tags based on standard words, is an easy-to-read format for humans while being straightforward for machines to process. EpIC assumes input data to be provided in units of GeV (and its powers, whenever applicable), and radians for angles. The general structure of a single scenario is the following:
\begin{xmlcode}
<!-- XML header -->
<?xml version="1.0" encoding="UTF-8" standalone="yes" ?>

<!-- Definition of scenario -->
<!-- For bookkeeping it includes date and description -->
<scenario date="2022-01-01" description="My first scenario">

<!-- Selection of service and its method-->
<task service="DVCSGeneratorService" method="generate">

  <!-- General configuration -->
  <general_configuration>
  ...
  </general_configuration>

  <!-- Selection of kinematic ranges -->
  <kinematic_range>
  ...
  </kinematic_range>

  <!-- Indication of experimental conditions -->
  <experimental_conditions>
  ...
  </experimental_conditions>
 
  <!-- Configuration of ProcessModule -->
  <computation_configuration>
  ...
  </computation_configuration>

  <!-- Configuration of GeneratorModule -->
  <generator_configuration>
  ...
  </generator_configuration>

  <!-- Configuration of KinematicModule -->
  <kinematic_configuration>
   ...
  </kinematic_configuration>

  <!-- Configuration of RCModule -->
  <rc_configuration>
  ...
   </rc_configuration>

  <!-- Configuration of WriterModule -->
  <writer_configuration>
  ...
  </writer_configuration>

</task>
</scenario>
\end{xmlcode}
Here, we only expose blocks containing specific information, like \xmlinline{<general_configuration> ... </general_configuration>}. Each block is described in one of the subsequent subsections. In our demonstration we only show an exemplary scenario for the DVCS case. More examples, including those for other processes, can be found online \cite{www:epic}.

\subsection{\texttt{general\_configuration}}

\begin{xmlcode}
<general_configuration>

  <!-- Number of events to be generated -->
  <param name="number_of_events" value="10" />
    
  <!-- Subprocess. Possible values: "DVCS" (pure DVCS), "BH" (pure Bethe-Heitler), "DVCS|BH" (pure DVCS and BH), "DVCS|BH|INT" (pure DVCS, BH and the interference between the two) -->
  <param name="subprocess_type" value="DVCS" />
  
</general_configuration>
\end{xmlcode}

\subsection{\texttt{kinematic\_range}}

\begin{xmlcode}
<kinematic_range>

  <!-- Range of $x_\text{B}$ (Bjorken variable)-->
  <param name="range_xB" value="0.|1." />
    
  <!-- Range of $t$ --> 
  <param name="range_t" value="-1.|0." />
    
  <!-- Range of $Q^{2}$ -->
  <param name="range_Q2" value="1.|10." />
    
  <!-- Range of $\phi$ -->
  <param name="range_phi" value="0.|2*pi" />
    
  <!-- Range of $\phi_{S}$ -->
  <param name="range_phiS" value="0.|2*pi" />
    
  <!-- Range of $y$ -->
  <param name="range_y" value="0.01|0.95" />
  
</kinematic_range>
\end{xmlcode}

\subsection{\texttt{experimental\_conditions}}

\begin{xmlcode}
<experimental_conditions>

  <!-- Energy of the lepton beam -->
  <param name="lepton_energy" value="5." />

  <!-- Type of the lepton beam, here electron -->
  <param name="lepton_type" value="e-" />
  
  <!-- Polarisation of the lepton beam -->
  <param name="lepton_helicity" value="1" />
  
  <!-- Energy of the hadron beam -->
  <!-- For a target in rest frame use: value="fixed\_target" -->
  <param name="hadron_energy" value="10." />
  
  <!-- Type of the hadron beam, here proton -->
  <param name="hadron_type" value="p" />
  
  <!-- Polarisation of the hadron beam -->
  <!-- Possible values:  "0|0|0" for unpolarised target, "0|0|$\pm 1$" for long. polarised target, "$\pm 1$|0|0" or "0|$\pm 1$|0" for trans. polarised target -->
  <param name="hadron_polarisation" value="0.|0.|0." />
  
</experimental_conditions>
\end{xmlcode}

\subsection{\texttt{computation\_configuration}}

\begin{xmlcode}
<computation_configuration>

  <!-- Selection of module -->
  <!-- DVCSProcessBMJ12 module encodes the BH, DVCS, and interference terms of the Born cross-section according to Refs. \cite{Belitsky:2001ns} and \cite{Belitsky:2012ch} -->
  <module type="DVCSProcessModule" name="DVCSProcessBMJ12">

    <!-- Selection of module used for evaluation of factorisation and renormalisation scales from DVCS kinematics -->
    <!-- DVCSScalesQ2Multiplier module identifies both scales in terms of the $Q^2$ variable -->
    <module type="DVCSScalesModule" name="DVCSScalesQ2Multiplier">
    </module>

    <!-- Selection of module used for evaluation of the GPD skewness variable, $\xi$, from DVCS kinematics -->
    <!-- DVCSXiConverterXBToXi module use the conversion $\xi = \xBj / (2 - \xBj)$ -->
    <module type="DVCSXiConverterModule" name="DVCSXiConverterXBToXi">
    </module>

    <!-- Selection of module used for the evaluation DVCS Compton form factors (CFFs) -->
    <!-- DVCSCFFCMILOU3DTables module does not evaluate CFFs from a GPD model during EpIC run-time. Instead, it uses look-up tables of CFFs evaluated ahead of time for a particular GPD module. A number of these tables are included with EpIC.-->
    <module type="DVCSConvolCoeffFunctionModule" name="DVCSCFFCMILOU3DTables">

      <!-- pQCD order of evaluation -->
      <param name="qcd_order_type" value="LO" />
  
      <!-- Path to a look-up table, here the one based on the GK GPD model is specified -->
      <param name="cff_set_file" value="PATH/epic/data/ DVCSCFFCMILOU3DTables/tables_GK.root" />
</module>
  
  </module>
</computation_configuration>
\end{xmlcode}

\subsection{\texttt{generator\_configuration}}

\begin{xmlcode}
<generator_configuration>

  <!-- Selection of module -->
  <!-- EventGeneratorFOAM module uses mini-FOAM library \cite{Jadach:2005ex} that is issued with ROOT \cite{Brun:1997pa} -->
  <module type="EventGeneratorModule" name="EventGeneratorFOAM">
  
    <!-- Parameters of the FOAM algorithm -->

    <!-- Maximum number of cells -->
    <param name="nCells" value="10000" />
    
    <!--Number of MC events when exploring a cell -->
    <param name="nSamples" value="2000" />
    
    <!-- Number of bins in edge histogram for a cell -->
    <param name="nBins" value="2000" />
    
    <!-- After the initialisation process, the state of FOAM will be saved in PATH/state.root. To use this file in another EpIC run, so as to skip the initialisation in that run, use the read\_state\_file\_path option. -->
    <param name="save_state_file_path" value="PATH/state.root" />
  </module>
</generator_configuration>
\end{xmlcode}

\subsection{\texttt{kinematic\_configuration}}

\begin{xmlcode}
<kinematic_configuration>

  <!-- Selection of module -->
  <!-- DVCSKinematicDefault module provides default evaluation of four-momenta from DVCS kinematics -->
  <module type="DVCSKinematicModule" name="DVCSKinematicDefault">
  </module>
</kinematic_configuration>
\end{xmlcode}

\subsection{\texttt{rc\_configuration}}

\begin{xmlcode}
<rc_configuration>

  <!-- Selection of module -->
  <!-- DVCSRCCollinear module provides evaluation of initial and final state radiative corrections from lepton lines using the collinear approximation, see Sect. \ref{sec:rc} for more details -->
  <module type="DVCSRCModule" name="DVCSRCCollinear">
  </module>
</rc_configuration>
\end{xmlcode}

\subsection{\texttt{writer\_configuration}}

\begin{xmlcode}
<writer_configuration>

  <!-- Selection of module -->
  <!-- WriterHepMC3 module is used to save event records in HepMC3 format \cite{Buckley:2019xhk} -->
  <module type="WriterModule" name="WriterHepMC3">
  
    <!-- Path to output file containing event records -->
    <param name="output_file_path" value="test.txt" />
    
    <!-- Format of output file containing event records: "ascii" for text format, or "root" for binary ROOT format-->
    <param name="HepMC3_writer_type" value="ascii" />
  </module>
</writer_configuration>
\end{xmlcode}

\section{Demonstration of EpIC}
\label{sec:demo}

Here, we demonstrate EpIC in a detailed manner by using specific components of the PARTONS framework. The demonstration is based on one million MC events generated for the DVCS sub-process, \ie exclusive leptoproduction of a single photon without the Bethe-Heitler contribution, at $10\,\mathrm{GeV}$ positive helicity electron and $100\,\mathrm{GeV}$ unpolarized proton energies. The following modules are used during the generation process: \texttt{DVCSCFFCMILOU3DTables} for the parameterisation of CFFs obtained from the GK GPD model \cite{Goloskokov:2005sd,Goloskokov:2007nt,Goloskokov:2009ia} and LO coefficients functions, \texttt{DVCSProcessBMJ12} for the evaluation of DVCS cross-section based on the set of expressions published in Ref. \cite{Belitsky:2012ch}, \texttt{DVCSRCNull} for the simulation without radiative corrections, \texttt{EventGeneratorFOAM} for the generation of DVCS kinematics based on cross-section with the FOAM algorithm \cite{Jadach:2005ex}, \texttt{DVCSKinematicDefault} for the default evaluation of four-momenta from DVCS kinematics, and \texttt{WriterHepMC3} to save four-momenta in HepMC3 format. 

\begin{figure*}[!ht]
\begin{center}
\includegraphics[width=1.0\textwidth]{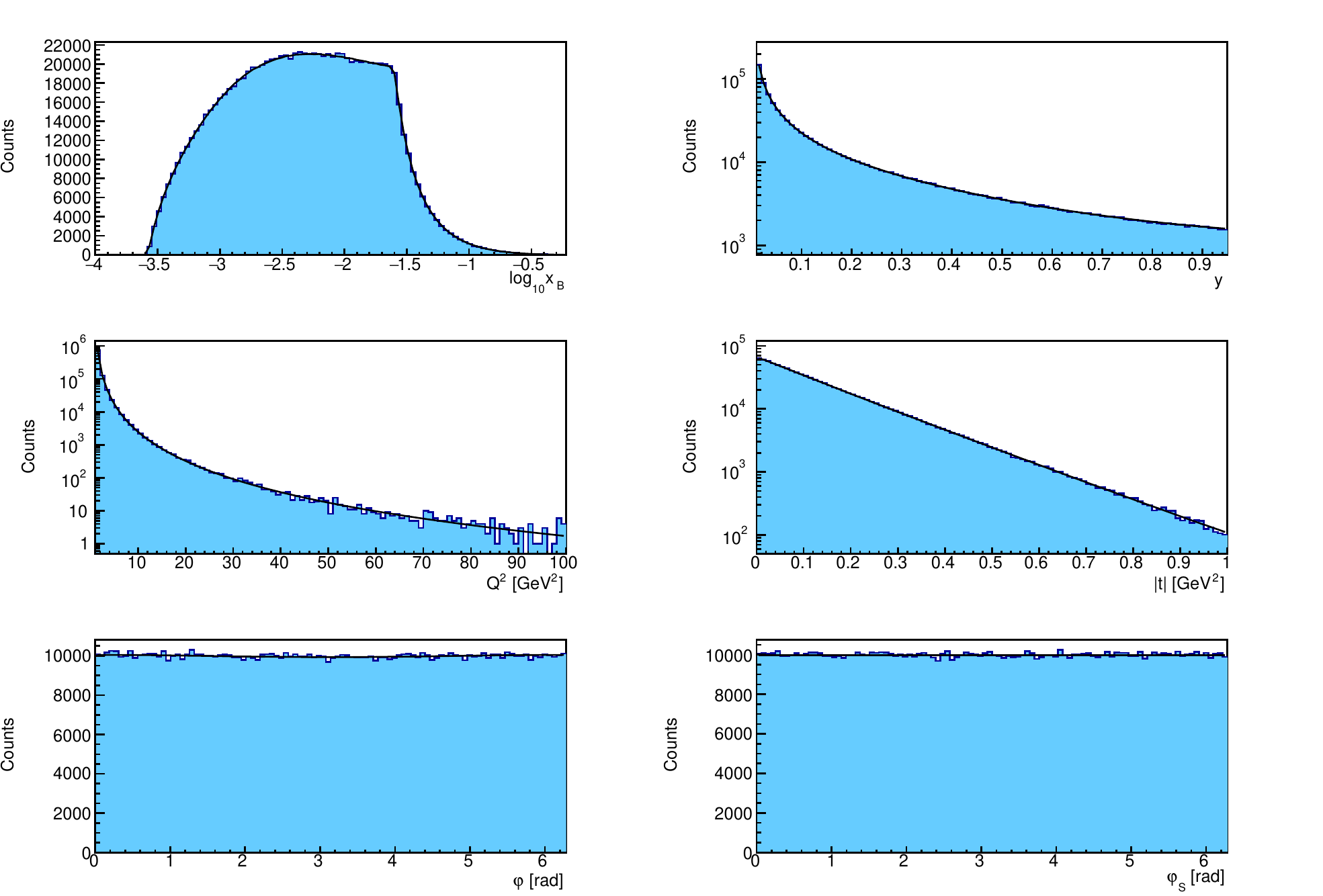}
\caption{Distribution of one million DVCS events produced by the \texttt{DVCSProcessBMJ12} module utilising \texttt{GPDGK16} GPDs within the PARTONS framework. Events are binned with respect to $\text{log}_{10}(x_\text{B}), y, Q^2, |t|, \phi$ and $\phi_\text{S}$. The solid curves on top of the histograms depict the expected number of events for the given model. The curves are normalised according to the integrated luminosity estimated by EpIC.}
\label{fig:demo}
\end{center}
\end{figure*}

The following cuts are used in the generation process: $0.0001\leq x_\text{B}\leq 0.6$, $0.01\leq y\leq 0.95$ (here, $y$ is the inelasticity parameter), $1\leq Q^2\leq 100$ $\text{GeV}^2$, $0\leq |t|\leq 1$ $\text{GeV}^2$, $0 \leq \phi \leq 2\pi$, and $0 \leq \phi_\text{S} \leq 2\pi$. On the basis of these cuts, the distribution of events is shown in Fig.~\ref{fig:demo}. The following FOAM library parameters are used in producing those events \cite{Jadach:2005ex}: $\texttt{nCells} = 3000$, $\texttt{nSamples} = 600$, $\texttt{nBins} = 600$. For the FOAM library, the initialisation lasted about $40$ minutes, whereas the generation time after initialisation took around $0.0052$ seconds per event at BNL computer farms. The initialisation time can be shortened by changing the parameters of the FOAM library, which may, however, lead to a longer time needed by EpIC to generate a single event. We remind that the result of the initialisation can be stored in a file, allowing to avoid this step in subsequent MC runs.

To ensure our results are consistent, we also plot the theory expectation values, as solid lines, on top of each histogram. The number of events associated with the theory curves is calculated as follows: at each bin, the cross-section for the given bin width is calculated, and then the result is divided by the total cross-section of the phase space. One can then multiply this ratio by the total number of events to find the number of events in that specific bin. To obtain those curves in Fig.~\ref{fig:demo}, first we assign the number of events to the middle point of each bin, then interpolate all these results. As a result, we observe that EpIC generates events in accordance with the theory values.

\section{Summary}
\label{sec:summary}

In this work, we described in detail the overall structure and essential features of the novel MC generator EpIC. This generator is based on the PARTONS framework in the model selection. It is designed to generate events for exclusive processes, such as DVCS, TCS, DVMP and more broadly to all exclusive processes implemented in the PARTONS framework. As these processes have the potential to reveal the 3D structure of the nucleon through GPDs, they have been studied extensively at facilities such as DESY, JLab, and CERN, and they will remain crucial for future EIC programmes.  

The EpIC MC generator stands out with its flexible architecture that utilises a modular programming paradigm. Consequently, the code is easier to understand, as each element of the architecture is designed to perform a specific task. The distribution of events generated by EpIC is highly precise since it utilises an accurate representation of the cross-section of the underlying process in the FOAM library. Besides generating accurate events, EpIC also allows users to generate events with radiative corrections as well as to run multi-channel analyses. On top of this, EpIC is released through GitHub as an open-source project under a GPL licence. In particular it means that users can trace code modifications and new contributors can easily join the development team. To the best of our knowledge, a tool such as EpIC is unique in the GPD community by its very nature. The comprehensive set of features and the variety of models available in EpIC make it ideal for the systematic study of the 3D structure of the nucleon. 

The generator was designed in a way that facilitates easy expansion. The encapsulation of elements is one of the fundamental characteristics of the generator when it comes to maintaining the project for a long time and with a changing team of developers. This allows us to easily adapt the architecture to the latest developments. Finally, to make this project as useful to the particle physics community as possible, we promote open-source standards and offer easy access to its development. This perspective is well suited to the experimental timeline of the 3D hadron structure community.

\begin{acknowledgements}
The authors thank Alexander Jentsch, Kre\v{s}imir Kumeri\v{c}ki, and Kornelija Passek-Kumeri\v{c}ki for fruitful discussions. 
This work was supported by the Grant No. 2019/35/D/ST2/00272 of the National Science Center, Poland; 
by the European Union's Horizon 2020 research and innovation programme under grant agreement No 824093; 
in the framework of the GLUODYNAMICS project funded by the "P2IO LabEx (ANR-10-LABX-0038)" in the framework "Investissements d’Avenir" (ANR-11-IDEX-0003-01) managed by the Agence Nationale de la Recherche (ANR), France; 
by the Ile-de-France region via the Blaise Pascal Chair of International Excellence.
KT was supported by U.S. Department of Energy under Contract No. de-sc0012704. 
\end{acknowledgements}

\appendix

\section{Radiative corrections}
\label{sec:rc}

Radiative corrections can have a significant impact on the interpretation of experimental data, as they introduce uncertainties in the reconstruction of ``true'' kinematics of the measured process. On the other hand, neglecting radiative corrections in simulations typically leads to unrealistic distributions of generated events, making \eg impact studies to be much less robust. To address both of these problems it is highly desirable to have radiative corrections implemented in generators of MC events.

In the literature, it has been discussed that the leptonic initial and final state radiative corrections, \ie when photons are emitted by either incoming or outgoing leptons, respectively, can be computed with a good accuracy in the collinear approximation \cite{Mo:1968cg,Kripfganz:1990vm}. In this approximation the transverse component of the momenta of emitted photons are neglected. Therefore, the radiated photons are restricted to move along the direction of the source leptons. 

The collinear approximation allows one to introduce a single parameter, $z_1$ or $z_3$, for the initial or final state radiation, respectively. The parameter describes the energy of the emitted photon:
\begin{equation}
    z_1 = \frac{E_e - E_\gamma}{E_e}\,, \qquad z_3 = \frac{E_{e^\prime}}{E_{e^\prime} + E_{\gamma^\prime}} \,,
\end{equation}
where $E_{e}$ ($E_{e^\prime}$) and $E_{\gamma}$ ($E_{\gamma^\prime}$) are the energies of the incoming (outgoing) lepton and the initial (final) state emitted photon, respectively. The DIS cross section can be then expressed by \cite{Kripfganz:1990vm}:
\begin{align}
    \frac{\intd^2\sigma}{\intd x \, \intd y} = 
    \int_{z_1^\text{min}}^1 \frac{\intd z_1}{z_1} D(z_1) \int_{z_3^\text{min}}^1 \frac{\intd z_3}{z_3^2} \overline{D}(z_3)\, \frac{y}{\hat{y}}\, \frac{\intd^2\hat{\sigma}_{\text{Born}}}{\intd\hat{x} \, \intd\hat{y}} \,,
    \label{dis_cross_section}
\end{align}
where $\intd^2\hat{\sigma}_{\text{Born}}$ is the differential Born cross-section evaluated for ``true'' variables, which are related
to the observed ones by \cite{Kripfganz:1990vm}:
\begin{align}
   \hat{x} = \frac{z_1 x y}{z_1 z_3 + y - 1} \,,~ \hat{y} = \frac{z_1 z_3 + y - 1}{z_1 z_3} \,,
\end{align}
and where due to kinematic limits one has:
\begin{equation}
    z_1^\text{min} = \frac{1-y}{1-x y}, \qquad z_3^\text{min} = 1-y(1-x)\,.
\end{equation}
In Eq.~\eqref{dis_cross_section} the radiator functions, $D(z_{1})$ and $\overline{D}(z_3)$, describe the weight of the rescaled cross-section in the presence of the initial and final state radiations, respectively. At the leading order $D(z_1)$ and $\overline{D}(z_3)$ are equal to each other and given by: 
\begin{align}
   D(z) & = \Bigg[ \delta(1-z)\Bigg[1+\frac{\alpha}{2\pi}L\Bigg(2\ln\,\epsilon + \frac{3}{2}\Bigg)\Bigg] \nonumber \\
   & + \theta(1-\epsilon-z)\frac{\alpha}{2\pi}L\frac{1+z^2}{1-z} \Bigg]\,,
   \label{radiator_function}
\end{align}
with $L = \ln(Q^2/m_l^2)$. Here, $\alpha$ is the fine-structure constant, and $m_l$ is the lepton mass. The parameter $\epsilon$ acts as a cutoff avoiding the generation of a very soft photons characterised by energies $E_\gamma \le \epsilon E_e$ and $E_{\gamma^\prime} \le \epsilon E_{e^\prime}$. It is crucial to choose the cutoff carefully so that $\epsilon E_e$ and $\epsilon E_{e^\prime}$ are much smaller than the energy resolution of the experiment. Due to the cutoff Eq.~\eqref{radiator_function} consists of two parts: the term proportional to the Dirac delta function collectively describes the contribution of very soft photons and virtual corrections, while the term proportional to the step function describes the effects of real radiation of photons with sizeable energies. 

The generalisation of leading order radiative corrections for the DVCS case can be achieved as follows:
\begin{align}
    \frac{\intd^5\sigma}{\intd x \, \intd Q^2 \, \intd t \, \intd \phi \, \intd \phi_S} & = \int_{z_1^\text{min}}^1 \intd z_1 z_1 D(z_1) \int_{z_3^\text{min}}^1 \frac{\intd z_3}{z_3^2} \overline{D}(z_3) \nonumber \\
    & \times \frac{y}{\hat{y}}\,  \frac{\intd^5\hat{\sigma}_\text{Born}}{\intd \hat{x} \, \intd \hat{Q}^2 \, \intd t \, \intd \, \phi \, \intd \phi_S}\,. 
\end{align}

\bibliographystyle{spphys}
\bibliography{bibliography}

\begin{thebibliography}{10}
\providecommand{\url}[1]{{#1}}
\providecommand{\urlprefix}{URL }
\expandafter\ifx\csname urlstyle\endcsname\relax
  \providecommand{\doi}[1]{DOI \discretionary{}{}{}#1}\else
  \providecommand{\doi}{DOI \discretionary{}{}{}\begingroup
  \urlstyle{rm}\Url}\fi

\bibitem{Muller:1994ses}
D.~M\"uller, D.~Robaschik, B.~Geyer, F.M. Dittes, J.~Ho\v{r}ej\v{s}i, Fortsch.
  Phys. \textbf{42}, 101 (1994).
\newblock \doi{10.1002/prop.2190420202}

\bibitem{Ji:1996ek}
X.D. Ji, Phys. Rev. Lett. \textbf{78}, 610 (1997).
\newblock \doi{10.1103/PhysRevLett.78.610}

\bibitem{Ji:1996nm}
X.D. Ji, Phys. Rev. D \textbf{55}, 7114 (1997).
\newblock \doi{10.1103/PhysRevD.55.7114}

\bibitem{Radyushkin:1996ru}
A.V. Radyushkin, Phys. Lett. B \textbf{385}, 333 (1996).
\newblock \doi{10.1016/0370-2693(96)00844-1}

\bibitem{Radyushkin:1997ki}
A.V. Radyushkin, Phys. Rev. D \textbf{56}, 5524 (1997).
\newblock \doi{10.1103/PhysRevD.56.5524}

\bibitem{Burkardt:2000za}
M.~Burkardt, Phys. Rev. D \textbf{62}, 071503 (2000).
\newblock \doi{10.1103/PhysRevD.62.071503}.
\newblock [Erratum: Phys.Rev.D 66, 119903 (2002)]

\bibitem{Burkardt:2002hr}
M.~Burkardt, Int. J. Mod. Phys. A \textbf{18}, 173 (2003).
\newblock \doi{10.1142/S0217751X03012370}

\bibitem{Burkardt:2004bv}
M.~Burkardt, Phys. Lett. B \textbf{595}, 245 (2004).
\newblock \doi{10.1016/j.physletb.2004.05.070}

\bibitem{Ji:1997gm}
X.D. Ji, W.~Melnitchouk, X.~Song, Phys. Rev. D \textbf{56}, 5511 (1997).
\newblock \doi{10.1103/PhysRevD.56.5511}

\bibitem{Polyakov:2002yz}
M.V. Polyakov, Phys. Lett. B \textbf{555}, 57 (2003).
\newblock \doi{10.1016/S0370-2693(03)00036-4}

\bibitem{Polyakov:2018zvc}
M.V. Polyakov, P.~Schweitzer, Int. J. Mod. Phys. A \textbf{33}(26), 1830025
  (2018).
\newblock \doi{10.1142/S0217751X18300259}

\bibitem{Lorce:2018egm}
C.~Lorc\'e, H.~Moutarde, A.P. Trawi\'nski, Eur. Phys. J. C \textbf{79}(1), 89
  (2019).
\newblock \doi{10.1140/epjc/s10052-019-6572-3}

\bibitem{Diehl:2003ny}
M.~Diehl, Phys. Rept. \textbf{388}, 41 (2003).
\newblock \doi{10.1016/j.physrep.2003.08.002}

\bibitem{Accardi:2012qut}
A.~Accardi, et~al., Eur. Phys. J. A \textbf{52}(9), 268 (2016).
\newblock \doi{10.1140/epja/i2016-16268-9}

\bibitem{AbdulKhalek:2021gbh}
R.~Abdul~Khalek, et~al., {Science Requirements and Detector Concepts for the
  Electron-Ion Collider: EIC Yellow Report} (2021).
\newblock {arXiv:physics.ins-det/2103.05419}

\bibitem{Anderle:2021wcy}
D.P. Anderle, et~al., Front. Phys. (Beijing) \textbf{16}(6), 64701 (2021).
\newblock \doi{10.1007/s11467-021-1062-0}

\bibitem{LHeCStudyGroup:2012zhm}
J.L. Abelleira~Fernandez, et~al., J. Phys. G \textbf{39}, 075001 (2012).
\newblock \doi{10.1088/0954-3899/39/7/075001}

\bibitem{Adams:2676885}
B.~Adams, et~al., {COMPASS++/AMBER: Proposal for Measurements at the M2 beam
  line of the CERN SPS Phase-1: 2022-2024}.
\newblock Tech. rep., CERN, Geneva (2019).
\newblock CERN-SPSC-2019-022, SPSC-P-360

\bibitem{Arrington:2021alx}
J.~Arrington, et~al., {Physics with CEBAF at 12 GeV and Future Opportunities}
  (2021).
\newblock {arXiv:nucl-ex/2112.00060}

\bibitem{Berthou:2015oaw}
B.~Berthou, et~al., Eur. Phys. J. C \textbf{78}(6), 478 (2018).
\newblock \doi{10.1140/epjc/s10052-018-5948-0}

\bibitem{Bacchetta:2004jz}
A.~Bacchetta, U.~D'Alesio, M.~Diehl, C.A. Miller, Phys. Rev. D \textbf{70},
  117504 (2004).
\newblock \doi{10.1103/PhysRevD.70.117504}

\bibitem{www:epic_github}
{EpIC} project on {GitHub}.
\newblock \url{https://github.com/pawelsznajder/epic}

\bibitem{www:epic}
{EpIC} project webpage.
\newblock \url{https://pawelsznajder.github.io/epic}

\bibitem{Grocholski:2019pqj}
O.~Grocholski, H.~Moutarde, B.~Pire, P.~Sznajder, J.~Wagner, Eur. Phys. J. C
  \textbf{80}(2), 171 (2020).
\newblock \doi{10.1140/epjc/s10052-020-7700-9}

\bibitem{Moutarde:2018kwr}
H.~Moutarde, P.~Sznajder, J.~Wagner, Eur. Phys. J. C \textbf{78}(11), 890
  (2018).
\newblock \doi{10.1140/epjc/s10052-018-6359-y}

\bibitem{Moutarde:2019tqa}
H.~Moutarde, P.~Sznajder, J.~Wagner, Eur. Phys. J. C \textbf{79}(7), 614
  (2019).
\newblock \doi{10.1140/epjc/s10052-019-7117-5}

\bibitem{Bertone:2021yyz}
V.~Bertone, H.~Dutrieux, C.~Mezrag, H.~Moutarde, P.~Sznajder, Phys. Rev. D
  \textbf{103}(11), 114019 (2021).
\newblock \doi{10.1103/PhysRevD.103.114019}

\bibitem{Chavez:2021llq}
J.M.M. Chavez, V.~Bertone, F.D.S. Borrero, M.~Defurne, C.~Mezrag, H.~Moutarde,
  J.~Rodr\'\i{}guez-Quintero, J.~Segovia, {Pion GPDs: A path toward
  phenomenology} (2021).
\newblock {arXiv:hep-ph/2110.06052}

\bibitem{Dutrieux:2021wll}
H.~Dutrieux, O.~Grocholski, H.~Moutarde, P.~Sznajder, Eur. Phys. J. C
  \textbf{82}(3), 252 (2022).
\newblock \doi{10.1140/epjc/s10052-022-10211-5}

\bibitem{Goloskokov:2005sd}
S.V. Goloskokov, P.~Kroll, Eur. Phys. J. C \textbf{42}, 281 (2005).
\newblock \doi{10.1140/epjc/s2005-02298-5}

\bibitem{Goloskokov:2007nt}
S.V. Goloskokov, P.~Kroll, Eur. Phys. J. C \textbf{53}, 367 (2008).
\newblock \doi{10.1140/epjc/s10052-007-0466-5}

\bibitem{Goloskokov:2009ia}
S.V. Goloskokov, P.~Kroll, Eur. Phys. J. C \textbf{65}, 137 (2010).
\newblock \doi{10.1140/epjc/s10052-009-1178-9}

\bibitem{www:cmake}
{CMake} project webpage.
\newblock \url{https://cmake.org}

\bibitem{Brun:1997pa}
R.~Brun, F.~Rademakers, Nucl. Instrum. Meth. A \textbf{389}, 81 (1997).
\newblock \doi{10.1016/S0168-9002(97)00048-X}

\bibitem{Buckley:2019xhk}
A.~Buckley, P.~Ilten, D.~Konstantinov, L.~L\"onnblad, J.~Monk, W.~Pokorski,
  T.~Przedzinski, A.~Verbytskyi, Comput. Phys. Commun. \textbf{260}, 107310
  (2021).
\newblock \doi{10.1016/j.cpc.2020.107310}

\bibitem{galassi2009gnu}
M.~Galassi, et~al., \emph{GNU Scientific Library: Reference Manual} (Network
  Theory, 2009).
\newblock ISBN 0954612078

\bibitem{Belitsky:2001ns}
A.V. Belitsky, D.~Mueller, A.~Kirchner, Nucl. Phys. B \textbf{629}, 323 (2002).
\newblock \doi{10.1016/S0550-3213(02)00144-X}

\bibitem{Belitsky:2012ch}
A.V. Belitsky, D.~M\"uller, Y.~Ji, Nucl. Phys. B \textbf{878}, 214 (2014).
\newblock \doi{10.1016/j.nuclphysb.2013.11.014}

\bibitem{Jadach:2005ex}
S.~Jadach, P.~Sawicki, Comput. Phys. Commun. \textbf{177}, 441 (2007).
\newblock \doi{10.1016/j.cpc.2007.02.112}

\bibitem{Mo:1968cg}
L.W. Mo, Y.S. Tsai, Rev. Mod. Phys. \textbf{41}, 205 (1969).
\newblock \doi{10.1103/RevModPhys.41.205}

\bibitem{Kripfganz:1990vm}
J.~Kripfganz, H.J. Mohring, H.~Spiesberger, Z. Phys. C \textbf{49}, 501 (1991).
\newblock \doi{10.1007/BF01549704}

\end{thebibliography}

\end{document}